\documentclass[preprint,showpacs]{revtex4}
\usepackage{graphicx,amsmath,amssymb,amsfonts,latexsym,color,dcolumn,bm}
\begin{document}

\title{Quantum Dynamical Effects as a Singular Perturbation for Observables\\
in Open Quasi-Classical Nonlinear Mesoscopic Systems}

\author{G.P. Berman}
\affiliation{Theoretical Division, MS B213, Los Alamos National
Laboratory, Los Alamos, NM 87545, USA}

\author{F. Borgonovi}
\affiliation{Dipartimento di Matematica e Fisica, Universit\'a
Cattolica, via Musei 41, 25121 Brescia, Italy}
\affiliation{INFN, Sezione di Pavia, Italy}

\author{D.A.R. Dalvit}
\affiliation{Theoretical Division, MS B213, Los Alamos National
Laboratory, Los Alamos, NM 87545, USA}

\date{\today}

\begin{abstract}
We review our results on a mathematical dynamical theory for
observables for open many-body quantum nonlinear bosonic systems for
a very general class of Hamiltonians. We show that non-quadratic
(nonlinear) terms in a Hamiltonian provide a singular ``quantum"
perturbation for observables in some ``mesoscopic" region of parameters.
In particular,  quantum effects result in secular terms in the dynamical
evolution, that grow in time. We argue that even
for open quantum nonlinear systems in the deep quasi-classical
region, these  quantum effects can survive after decoherence and
relaxation processes take place. We demonstrate that these quantum
effects in open quantum systems can be observed, for example, in the
frequency Fourier spectrum of the dynamical observables, or in
the corresponding spectral density of noise. Estimates are
presented for Bose-Einstein condensates, low temperature mechanical
resonators, and nonlinear optical systems prepared in
large amplitude coherent states. In particular, we show that for Bose-Einstein condensate
systems the characteristic time of deviation of quantum dynamics for
observables from the corresponding classical dynamics coincides with
the characteristic time-scale of the well-known quantum nonlinear
effect of phase diffusion.
\end{abstract}


 \maketitle

\section{Introduction}

Real physical systems are not isolated, they are coupled to external
degrees of freedom. The classical and quantum dynamics of these open
systems are especially complex for nonlinear systems
that exhibit several phenomena,
including deviation of quantum dynamics from the corresponding
classical one, quantum revivals, decoherence, and relaxation.
Recently substantial effort has been devoted to study the open
dynamics of nonlinear quantum systems, with the aim of understanding
the quantum to classical transition in a controlled way \cite{nl}

Standard mathematical treatments of open quantum nonlinear systems
suffer from problems arising from the interplay between the
nonlinearity and the openness of the system. Usually the dynamics of
open quantum systems is studied using different mathematical
approaches, such as the master equation for the reduced density
matrix, which is an average of the full density matrix over the environment.
\cite{Giulini,ZurekRMP,Paz2001},
and quasi-probability distributions (e.g. the so-called Q-function \cite{Gardiner2000},
the Wigner function \cite{Agarval1970}, etc).
Although all of these approaches allow one, in principle, to calculate the time evolution of the
average values of the dynamical variables of the system, they have
significant drawbacks. In particular, these distribution functions
may not be positively defined; they may be inconsistent for certain
density matrices; it may be difficult to extract physical
information from these distributions, especially in the context of
quantum nonlinear open systems; in the ``deep" quasi-classical
region of parameters, $\epsilon = \hbar / J \ll 1$ (where  $\hbar$
is Planck constant and $J$ is a characteristic action of the
corresponding classical system) these quasi-probability
distributions exhibit fast oscillations due to phases like
$\exp(i S(t)/ \hbar)$, with $|S(t)| \simeq J$. Therefore, it is difficult to
separate the physical effects for dynamical observables
(requiring an additional multi-dimensional integration of quasi-distribution
densities)  from the effects  of errors related to a concrete mathematical approach.

We are approaching these problems using an alternative strategy that
starts from a mathematical dynamical theory based on \emph{exact},
linear partial differential equations (PDEs) for the observables of
open many-body quantum nonlinear bosonic systems governed by a very
general class of Hamiltonians (see \cite{Berman1994,Berman2004,Dalvit2006}
and references therein). The key advantage of this method is that it leads to a well-behaved
asymptotic theory for open quantum systems in the quasi-classical
region of parameters.  This approach is a generalization to the open
case of the asymptotic theory for bosonic and spin closed quantum
systems \cite{Berman1994,Vishik2003,Berman2003}, and it can be
applied to general open quantum nonlinear bosonic and spin systems
for a large range of parameters, including the deep quasi-classical
region.


We concentrate our attention on a discussion of the method which can
be used to observe quantum effects after decoherence and relaxation,
in the deep quasi-classical region of parameters. We argue that one
can use for these purposes a Fourier spectrum of the dynamical
observables, since its width contains characteristic information of
such quantum effects. Our observation is based on our first studies
\cite{Dalvit2006,Berman2004} of this new approach to quantum
nonlinear systems interacting with an environment. As will be
discussed below, certain quantum effects which are presented in the
dynamics of these nonlinear systems are robust to the influence of
the environment, and survive after decoherence and relaxation
processes take place.  In order to observe these effects
experimentally it is necessary to have a quasi-classical system in
certain region of parameters. We call these systems ``mesoscopic",
mainly because the parameter $\epsilon$ should not be too small.
In this sense, many quasi-classical systems have  the drawback that
they are either ``too classical" (i.e., they have a large $J$ so
that the quasi-classical parameter $\epsilon$ is extremely small),
or they interact too strongly with the environment, or their
effective temperature is so high that quantum effects that we are
talking about are washed out. Only recently have adequate open
nonlinear quasi-classical systems become available, including
Bose-Einstein condensates (BEC) with large number of atoms and
thermally well isolated; high frequency cantilevers with large nonlinearities
and at sufficiently low temperatures; and nonlinear optical systems in high Q resonators,
among others. We present estimates on the parameter regions where
survival of certain quantum effects to environment-induced
decoherence can be observed in these systems.


\section{Dynamics of quantum observables for closed quantum nonlinear systems}

We first consider closed quantum nonlinear systems. As a
simple example we take the one-dimensional quantum nonlinear
oscillator (QNO) described by the Hamiltonian
\cite{BermIomiZasl1,Berman1994} (see also an application of this
Hamiltonian for the BEC system in Section V)
\begin{equation}
H_s = \hbar \omega a^{\dagger} a + \mu \hbar^2 (a^{\dagger} a)^2 \;
, \ [a^{\dagger},a]=1 ,
\label{QNO}
\end{equation}
where  $a, a^{\dagger}$ are the annihilation and creation operators,
$\omega$ is the frequency of linear oscillations, and  $\mu$ is a
dimensional parameter of nonlinearity. We assume that initially the
QNO is prepared in a coherent state $|\alpha \rangle \; (a |\alpha
\rangle = \alpha |\alpha \rangle$). In the classical limit ($a
\rightarrow \alpha, a^{\dagger} \rightarrow \alpha^*, |\alpha|^2
\rightarrow \infty, \hbar |\alpha|^2 = J$, the classical action of
the linear oscillator) the Hamiltonian (\ref{QNO}) becomes $H_{\rm
cl} = \omega J + \mu J^2$. Below  we use the following dimensionless
notation:  $\tau \equiv \omega t$, $\bar{\mu} \equiv \hbar \mu/
\omega$, and $\mu_{\rm cl} \equiv \mu J/ \omega$. The quantum
parameter of nonlinearity $\bar{\mu}$ can be presented as the
product of two parameters, quantum and classical, $\bar{\mu}=
\epsilon \mu_{\rm cl}$. The parameter  $\mu_{\rm cl}$ characterizes
the nonlinearity in the classical nonlinear oscillator (BEC,
cantilever, optical field, etc) and can be written as   $\mu_{\rm
cl} = (J/2 \omega) (d\omega_{\rm cl} / dJ$), where $\omega_{\rm cl}
= dH_{\rm cl} / dJ = \omega + 2 \mu J$  is the classical frequency
of nonlinear oscillations. The limit  $\mu_{\rm cl} \ll 1$
corresponds to weak nonlinearity, while $\mu_{\rm cl} \simeq 1$
corresponds to strong nonlinearity. As was mentioned above,
$\epsilon$  is the quasi-classical parameter. Namely, $\epsilon
\simeq 1$ corresponds to the pure quantum system, and $\epsilon \ll
1$ corresponds to the quasi-classical limit, which is the subject of
our interest.

\subsection{Closed partial differential equation for observables}

\begin{figure}[t]
\begin{center}
\includegraphics[scale=0.4]{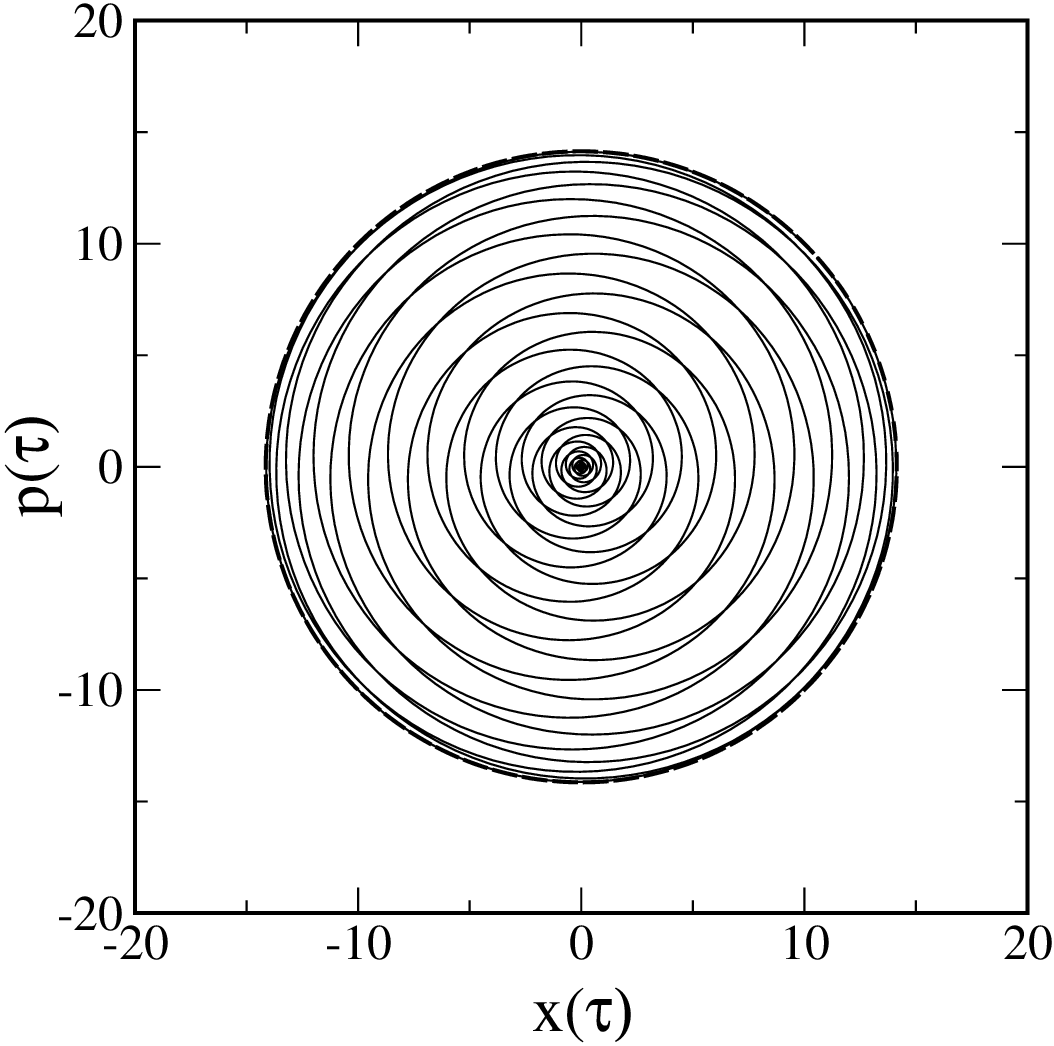}
\end{center}
\caption{Quasi-classical dynamics as described by the
observable in Eq.~(\ref{solutionclosed}).
Parameters are
$\epsilon=0.02$, $\bar{\mu} = 0.01$, $\tau_{\hbar} = 5$, $\tau_{\rm R}=314$,
$\tau_{\rm cl}=2.09$, $|\alpha|^2=100$, $\mu_{\rm cl}=1$.
Hence $\tau_{\rm cl} < \tau_{\hbar} < \tau_{\rm R}$.}
\label{uno}
\end{figure}

A closed linear PDE which describes the time evolution of the
expectation value of any observable of the system can be easily derived
when the system is initially populated in a coherent state $|\alpha
\rangle$ (see \cite{Berman1994} and references therein). Namely, for
an arbitrary operator function $f=f(a,a^{\dagger})$, the
time-dependent expectation value (observable) of such a function,
\begin{equation}
f(\alpha^*,\alpha,\tau)=\langle \alpha | e^{ i H t/ \hbar} f e^{-i H t/\hbar} | \alpha \rangle ,
\end{equation}
satisfies a PDE of the form
\begin{equation}
\partial f  / \partial \tau = \hat{K} f,
\end{equation}
where$\hat{K} = \hat{K}_{\rm cl} + \epsilon \mu_{\rm cl} \hat{K}_{\rm q}$.
Here the operator $\hat{K}_{\rm cl}$  includes only the first
order derivatives and describes the corresponding classical limit,
while the operator  $\hat{K}_{\rm q}$ includes higher-order
derivatives and contains the quantum effects. For the Hamiltonian
(\ref{QNO}) we have
\begin{eqnarray}
\frac{\partial f}{\partial \tau} &=& i (1 +\bar{\mu} + 2 \bar{\mu} |\alpha|^2)
\left(
\alpha^* \frac{\partial}{\partial \alpha^*} -
\alpha \frac{\partial}{\partial \alpha}
\right)
\nonumber \\
&& + i \bar{\mu}
\left(
(\alpha^*)^2 \frac{\partial^2}{\partial(\alpha^*)^2} -
\alpha^2 \frac{\partial^2}{\partial \alpha^2}
\right) f .
\label{PDEclosed}
\end{eqnarray}
In particular, for the operator function $f(\tau=0)=a$, the evolution
of  $f(\tau)$ corresponds to the evolution of $\alpha(\tau) =
\langle \alpha | a(\tau) | \alpha \rangle$, with the initial
condition $\alpha(\tau=0) = \alpha$. In this case Eq.
(\ref{PDEclosed}) can be solved exactly
\cite{Berman1994,BermIomiZasl1}
\begin{equation}
\alpha(\tau) = \alpha \, e^{-i (1+ \bar{\mu})\tau} \; e^{|\alpha|^2 (
e^{-2 i \bar{\mu} \tau} -1)} .
\label{solutionclosed}
\end{equation}
Fig. \ref{uno} depicts the dynamics described by the observable in
Eq.~(\ref{solutionclosed})  in the coordinate-momentum plane. The
effective coordinate is defined as
$x(\tau) = (\alpha^*(\tau) + \alpha(\tau))/\sqrt{2}$,
and the effective momentum is defined as
$p(\tau) = i (\alpha^*(\tau) - \alpha(\tau)) / \sqrt{2}$.
The corresponding classical dynamics is described by the function
$\alpha_{\rm cl}(\tau) = \alpha e^{-i (1+2 \mu_{\rm cl}) \tau}$,
which corresponds to the circumference in Fig.~\ref{uno}.
Note that Eq. (\ref{PDEclosed}) maintains its form for any
observable $f$, but the initial conditions for different
observables are different. This is also the case for any quantum nonlinear
Hamiltonian with many degrees of freedom.


\subsection{Characteristic time-scales for a closed quantum nonlinear system}

The solution (\ref{solutionclosed}) has three characteristic
time-scales \cite{BermIomiZasl1,Berman1994,Berman2002,Pit1}.
In the limit $\bar{\mu} \tau \ll 1$, it can be re-written in the form
\begin{equation}
\alpha(\tau) = \alpha_{\rm cl}(\tau) e^{- \tau^2/2 \tau_{\hbar}^2}
\; \left[ 1+ O(\bar{\mu} \tau) + O(|\alpha|^2 \bar{\mu}^3 \tau^3)
\right] .
\label{solutionapprox}
\end{equation}
The first time-scale is the characteristic classical time-scale,
which can be chosen as the period of classical nonlinear oscillations,
\begin{equation}
\tau_{\rm cl} = \frac{2 \pi \omega}{ \omega_{\rm cl}} = \frac{2 \pi }{1+ 2 \mu_{\rm cl}} .
\end{equation}
The second time-scale is a
characteristic time of departure of the quantum dynamics from the
corresponding classical one
\begin{equation}
\tau_{\hbar} = \frac{1}{ 2 \bar{\mu}  |\alpha|} .
\label{time_hbar}
\end{equation}
This time-scale characterizes the departure of
quantum dynamics from the classical one for classically stable
systems. Historically, this time-scale  was introduced for classically unstable
(classically chaotic) systems in \cite{Berman1978}, and it was shown to have a
logarithmic dependence on $\epsilon$ (see also
\cite{Chirikov1991,Chirikov1988}).  The time-scale $\tau_{\hbar}$
is usually called the Ehrenfest time.
The amplitudes of quantum and classical observables coincide at
multiple times of the quantum recurrence time-scale, which is the
third characteristic time-scale,
\begin{equation}
\tau_{\rm R} = \frac{\pi}{ \bar{\mu} } .
\end{equation}

Since we are interested in the quasi-classical region of parameters,
it is reasonable to impose the following inequalities on these three
characteristic time-scales: $\tau_{\rm cl} < \tau_{\hbar} \ll
\tau_{\rm R}$. In our case, $\tau_{\rm cl} / \tau_\hbar = 4 \pi
\bar{\mu} |\alpha| / (1 + 2 \mu_{\rm cl}) \approx \pi
\sqrt{\epsilon} \ll 1$, and $\tau_{\hbar}/ \tau_{\rm R} \approx
\sqrt{\epsilon}/\pi \ll 1$. When deriving the first inequality, we
used the conditions $|\alpha|^2 \simeq J/\hbar = 1/\epsilon$ and
$\mu_{\rm cl} \simeq 1$, which corresponds to the condition of
strong nonlinearity. Note that the condition $|\alpha|^2 \bar{\mu}^3
\tau^3 \simeq 1$ (see the third term in (\ref{solutionapprox}) in
the square brackets in the expression for $\alpha(\tau)$) gives the
characteristic times $\tau \gg \tau_{\hbar}$, namely $\tau /
\tau_{\hbar} = 2 / \epsilon^{1/6} \gg 1$. This means that the third
term in Eq.~(\ref{solutionapprox}) is small on the time scale
$\tau_{\hbar}$. For the values of parameters in Fig.~\ref{uno} the
inequalities $\tau_{\rm cl} < \tau_{\hbar} \ll \tau_{\rm R}$ are
satisfied.

\subsection{Quantum effects as a singular perturbation to the
classical solution}

As was mentioned above, the  form of the differential operator
$\hat{K}$ is
$$\hat{K} = \hat{K}_{\rm cl} + \epsilon \mu_{\rm cl} \hat{K}_{\rm q}.$$
The operator $\hat{K}_{\rm cl}$   includes only the first order
derivatives and describes the classical dynamics of the system.
Usually, the corresponding classical solution can be found by the
method of characteristics, or some alternative well-developed
methods. Note that even this part of the solution can be rather
complicated, especially for classically unstable and chaotic
systems, and usually requires large-scale numerical simulations.
(See details for closed quantum nonlinear systems and quantum
nonlinear systems interacting with the time-periodic fields
\cite{Berman1994}.) Another example which demonstrates the
application of the approach based on PDEs with the operator
$\hat K$ is considered in \cite{Berman2002} for a unstable quantum
nonlinear system describing the dynamics of a Bose-Einstein
condensate with attractive interactions.

For quantum linear systems ($\mu_{\rm cl}=0$) the quantum effects
vanish  for any values of the quasi-classical parameter $\epsilon$.
The differential operator $\hat{K}_{\rm q}$ includes second and
higher order derivatives, and it describes quantum effects. The
solutions of these PDEs are well behaved in the quasi-classical
region, $\epsilon \ll 1$,   and in contrast to the fast oscillating
WKB solutions (typical of standard methods based on
quasi-probability distributions), our method leads to the so-called
Laplace-type expansions \cite{Vishik2003}. The crucial property of
the Laplace asymptotics is that the dynamical observables are
exponentially localized in phase space around coherent states.

Quantum effects for observables represent a singular perturbation to the classical
solution. Indeed, in the quasi-classical region, quantum terms in
the PDEs are represented by the product of the small parameter
$\epsilon$ times high order derivatives. Consequently, these quantum
terms lead to a secular behavior of the solution, which diverges in
time from the corresponding classical solution.
Only the case $\epsilon=0$ (for finite $\mu_{\rm cl}$) corresponds
to the exact classical limit. But the problem with this limit is
that for any real system $\epsilon \neq 0$   (because $\hbar \neq 0$
and $J \neq \infty$). Then, even a very small value of $\epsilon$
still ``mathematically" results in a singular perturbation to the
classical solution due to the quantum terms.

The singularity arising from the quantum terms reminds, up to some extent,
of the singularity provided by a ``small'' viscosity in the
Navier-Stokes (NS) equation, describing the dynamics of liquid and gas
flows. Indeed, in the NS equation a small viscosity multiplies the
higher order spatial derivatives. Then, even for very large Reynolds
numbers (when the nonlinear terms are very large compared to the viscous ones),
the viscosity plays a crucial role in the dynamics of the flow,
even though it formally represents a ``small" perturbation.
Similarly, in the quantum case the small parameter $\epsilon$ multiplies the higher order
derivatives, which results in a quantum singular perturbation
for observables  even in the ``deep'' quasi-classical region.
It is this singularity that leads to a significant difference from the
classical solution.


\subsection{Frequency Fourier spectrum for quantum observables}


\begin{figure}[t]
\begin{center}
\includegraphics[scale=0.35]{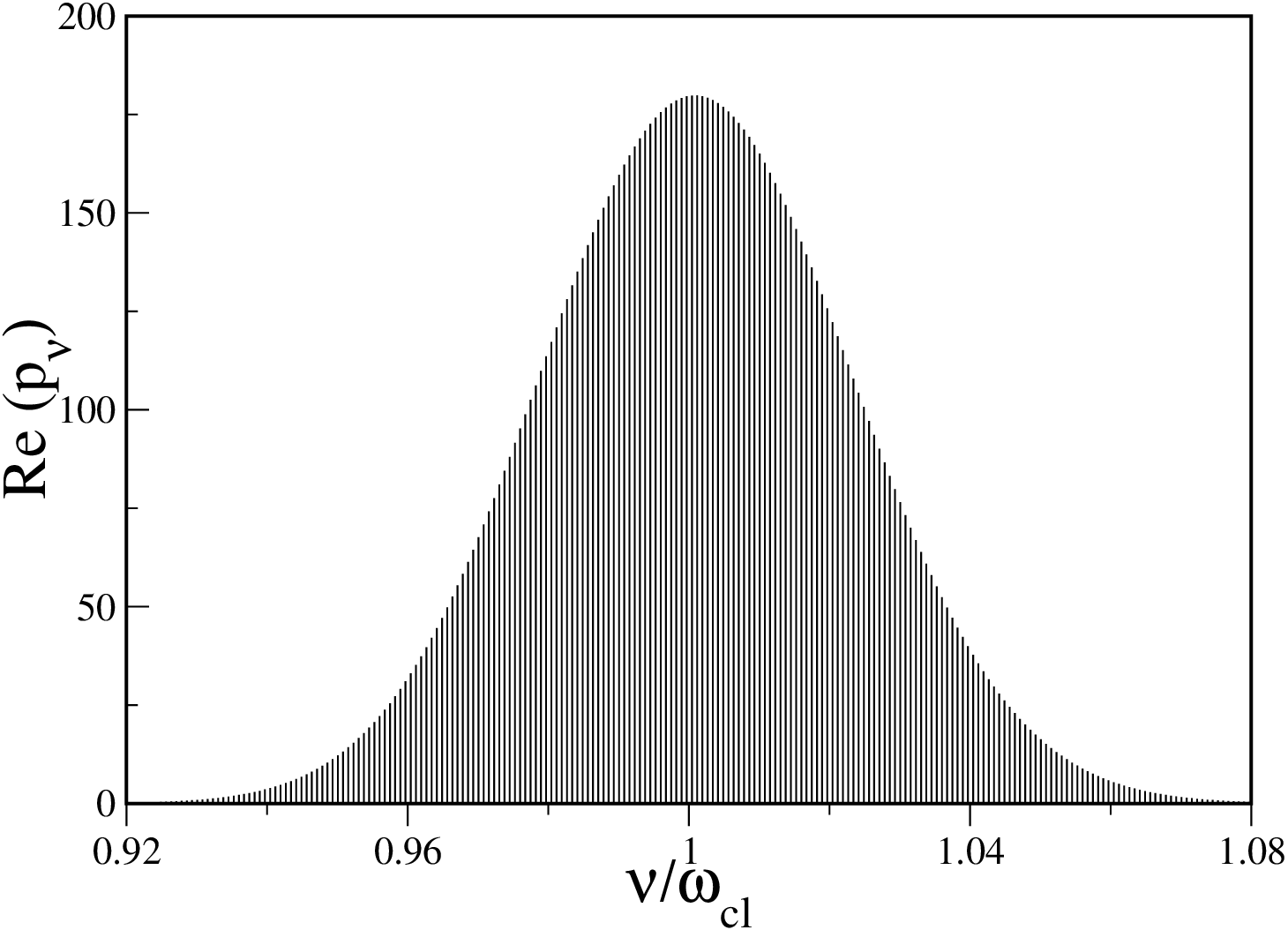}
\end{center}
\caption{Frequency spectrum of the effective momentum $p(\tau)$.
Parameters are
$\epsilon=1/900$, $\bar{\mu} =1/900$,
$\tau_{\hbar} \approx 15$, $\tau_{\rm R} \approx 900 \pi$,
$\tau_{\rm cl} \approx 2 \pi /3$, and $|\alpha|^2=900$.
Hence $\tau_{\rm cl} < \tau_{\hbar} \ll \tau_{\rm R}$.
}
\label{due}
\end{figure}

The observable $\alpha(\tau)$ can be written in the form
\begin{equation}
\alpha(\tau) = \alpha e^{-i (1+\bar{\mu}) \tau -
i |\alpha|^2 \sin(2 \bar{\mu} \tau)} \;
e^{-2 |\alpha|^2 \sin^2(\bar{\mu} \tau)} .
\label{five}
\end{equation}
The first exponent in Eq. (\ref{five}) is responsible for phase
modulations of the classical dynamics, while the second one is
responsible for amplitude modulations. The characteristic time-scale
of the amplitude modulations, $\tau_{\rm am}$,  is defined by the
condition $|\alpha|^2 \bar{\mu}^2 \tau_{\rm am}^2 \approx 1$, or by
the time-scale $\tau_{\rm am} \approx \tau_{\hbar}$.  The time-scale
of phase modulations of the classical dynamics is defined by the
condition $|\alpha|^2 \bar{\mu} \tau_{\rm ph} \approx 1$, or
$\tau_{\rm ph} \approx \tau_{\hbar} / \sqrt{\epsilon} \gg
\tau_{\hbar}$. Thus, the shortest time-scale which characterizes the
deviation of the quantum dynamics from the corresponding classical
one is the time $\tau_\hbar$. Moreover, this time-scale is
responsible for the finite  width of the spectral line $\Delta
\nu_{\hbar} \approx 2 \sqrt{2} / \tau_{\hbar}$.

Fig. \ref{due} depicts the frequency Fourier spectrum of the
effective momentum $p(\tau)$, with initial condition $p(0)=0$. One
can see that the frequency spectrum consists of one central line
with $\nu = \omega_{\rm cl} = 1+ 2 \mu_{\rm cl}$, and a width which
is approximately equal to $\Delta \nu \approx \Delta \nu_{\hbar}$.
In our case  the analytical estimate gives $\Delta \nu_{\hbar}
\approx 2 \sqrt{2} / \tau_{\hbar} \approx 0.19$,  which is very
close to the numerical results presented in Fig.~\ref{due}, $\Delta
\nu \approx 0.183$. The fine structure of the frequency spectrum is
provided by the characteristic revival time scale $\tau_{\rm R} =
\pi / \bar{\mu}$, or by the frequencies $\nu_n = 2 \bar{\mu} n$,
which are responsible for the complicated dynamics of quantum
recurrences.


\section{Dynamics of quantum observables for open quantum nonlinear systems}

The Hamiltonian of  open quantum nonlinear system interacting with
an environment contains three terms,
\begin{equation}
\hat{H} = \hat{H}_{\rm  S} + \hat{H}_{\cal E} + \hat{H}_{\rm int} .
\end{equation}
The first term is typically a time-independent polynomial
Hamiltonian of a general form which describes the self evolution
of the closed system,
$$\hat{H}_{\rm S} = \sum_{l,s} H_{l,s}
a^{\dagger l_1}_1 \ldots a^{\dagger l_N}_{N}
a^{s_1}_1 \ldots a^{s_N}_N,$$ where  $H_{l,s} = H^*_{l,s}$,
$l=(l_1,\ldots,l_N) \in Z^N_+$, and $s=(s_1,\ldots,s_N) \in Z^N_+.$
The operators $a_l$ and $a^{\dagger}_k$  satisfy bosonic
commutation relations,
$[a_l,a^{\dagger}_k] = \delta_{l,k}$.
A particular system corresponds to a particular choice of the
coefficients $H_{l,s}$ in $\hat{H}_{\rm S}$. The second term is
the Hamiltonian of the environment, which, for example, can be
modeled by a collection of harmonic oscillators,
\begin{equation}
\hat{H}_{\cal E} = \sum_{\vec q} \hbar \omega_{\vec q} \, b^{\dagger}_{\vec q} \, b_{\vec q} .
\label{qbm}
\end{equation}
Usually the oscillators of the environment are assumed to be
initially in thermal equilibrium,
$$\rho_{\cal E}(t=0) = Z_{\cal E}^{-1}
e^{-\hat{H}_{\cal E} / k_{\rm B} T},$$
where $Z_{\cal E} = {\rm Tr} [ e^{-\hat{H}_{\cal E} / k_{\rm B} T} ]$
is the partition function of the environment, $T$ is the temperature of the
environment,  and $k_{\rm B}$ is Boltzmann constant.
The third term is the interaction Hamiltonian between the system and the
environment. Prototype examples are the dipole-dipole interaction
Hamiltonian,
\begin{equation}
\hat{H}_{\rm int} = \hbar\sum_{n,{\vec q}} \lambda_{n,{\vec q}}
[ (a^{\dagger}_n + a_n) (b^{\dagger}_{\vec q} + b_{\vec q}) ] ,
\label{dipole_dipole}
\end{equation}
and the density-density interaction Hamiltonian
\begin{equation}
\hat{H}_{\rm int} = \hbar^2a^{\dagger} a \sum_{\vec
q}\lambda_{\vec q} \, b^{\dagger}_{\vec q} \, b_{\vec q} .
\label{density_density}
\end{equation}

\subsection{The differential operator $\hat{K}$  for many-body systems}

In a general many-body system the differential
operator $\hat{K}$ can formally be written as
\begin{eqnarray}
\hat{K} &=& \frac{i}{\hbar} e^{-\sum (|\alpha_n|^2 + |\beta_{\vec
q}|^2)} \sum \left[ H\left( \alpha^*_l, \beta^*_q,
\frac{\partial}{\partial \alpha^*_l}, \frac{\partial}{\partial
\beta^*_q}
\right) \right. \nonumber \\
&& \left. -  H\left( \alpha_l, \beta_q, \frac{\partial}{\partial
\alpha_l},   \frac{\partial}{\partial \beta_q} \right) \right] \;
e^{-\sum (|\alpha_n|^2 + |\beta_{\vec q}|^2)}  .
\end{eqnarray}
Note that after explicit differentiations, exponents in $\hat{K}$
vanish. Specific examples considered in our previous works include:
(i) a closed quantum one-dimensional nonlinear system in the
vicinity of an elliptic  \cite{Berman1994,Berman2003} or a hyperbolic
\cite{Berman2003, Berman2002} point;  (ii) chaotic systems describing the
interaction of atoms with radiation and external radio frequency fields \cite{Berman1994};
and (iii) the quantum Brownian motion problem for a nonlinear system oscillator
\cite{Dalvit2006,Berman2004}.

\subsection{Frequency Fourier spectrum of $p(\tau)$  in the presence of an environment}

\begin{figure}[t]
\begin{center}
\includegraphics[scale=0.35]{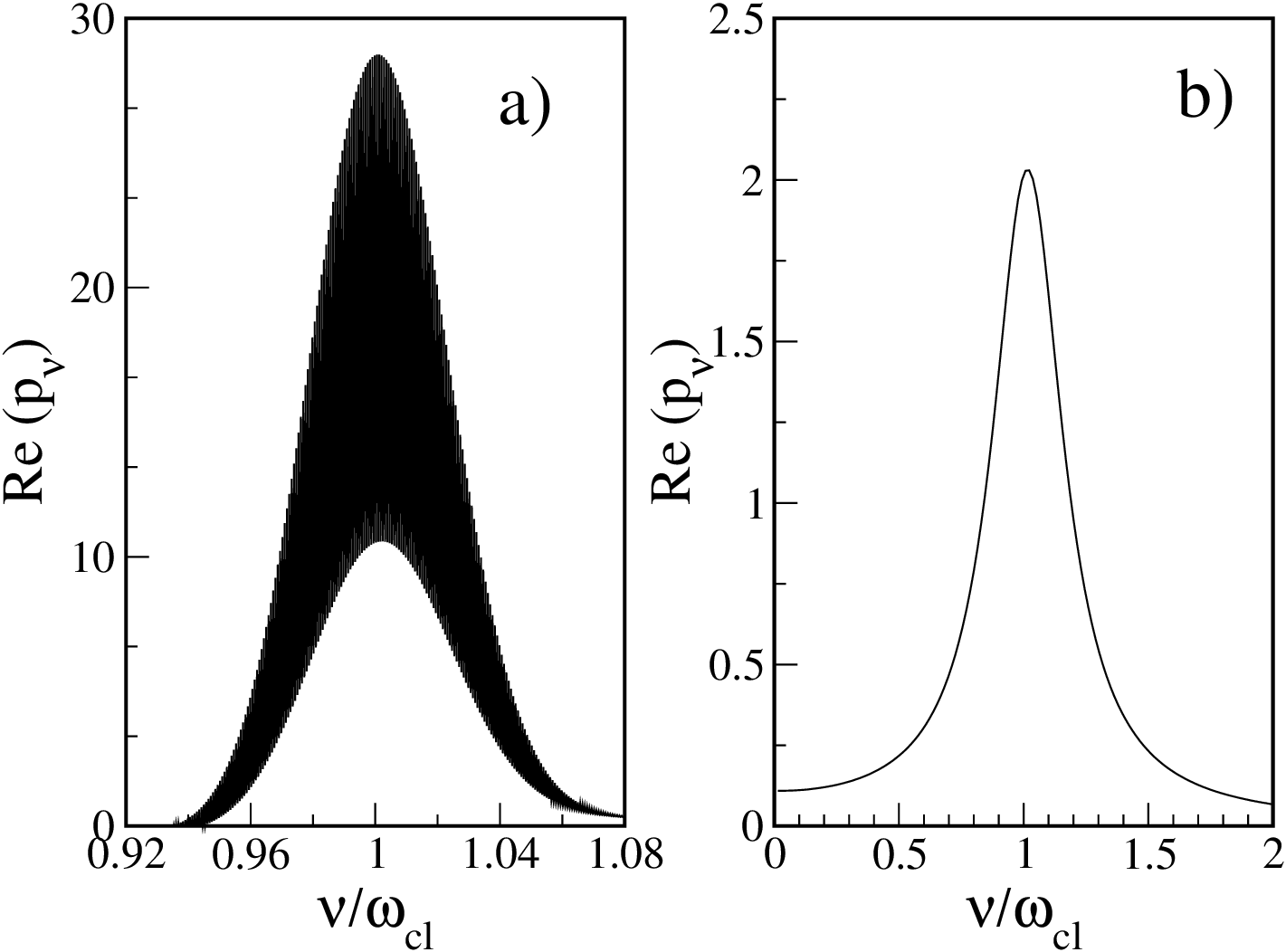}
\end{center}
\caption{Fourier frequency spectrum of the momentum $p(\tau)$
obtained from Eq.~(\ref{solutionopen}).
Parameters are: a) $\gamma=0.0005$, $\tau_{\gamma} =2000 \gg \tau_{\rm \hbar}
\approx 15$; b) $\gamma=0.5$,
$\tau_{\gamma} = 2  < \tau_{\hbar} \approx 15$; all other
parameters are the same as in Fig.~\ref{due}.}
\label{tre}
\end{figure}

Let us introduce formally a relaxation (dissipation) term into Eq.~(\ref{solutionclosed}).
Namely, we consider the function
\begin{equation}
\alpha(\tau) = \alpha \,  e^{-\gamma \tau -i
(1+\bar{\mu}) \tau} \; e^{|\alpha|^2 (e^{-2 i \bar{\mu} \tau} -1)} ,
\label{solutionopen}
\end{equation}
where the parameter $\gamma$  plays the role of an effective relaxation.
The characteristic time scale of relaxation is
$\tau_{\gamma} = 1/\gamma$.  We consider the frequency Fourier spectrum
of the momentum
$p(\tau) = i (\alpha^*(\tau) - \alpha(\tau))/\sqrt{2},$
with $\alpha(\tau)$ given by Eq.~(\ref{solutionopen}), for two cases:
(i)  $\tau_{\gamma} \gg \tau_{\hbar}$ (Fig.~\ref{tre}a), and (ii)
$\tau_{\gamma} < \tau_{\hbar}$  (Fig.~\ref{tre}b) (similar
dependencies can be built for the effective coordinate $x(\tau)$).
As one can see, when the influence of the effective dissipation is
small (Fig.~\ref{tre}a), the width of the Gaussian spectral line (at
the level $e^{-1}$) is still determined by the  time-scale
$\tau_\hbar$ ($\Delta \nu_{\hbar} \simeq 2 \sqrt{2} / \tau_{\hbar}
\approx 0.19$), and not by the environment ($\Delta \nu_{\gamma}
\simeq 2 \gamma =0.001$). The numerical results give $\Delta \nu
\approx 0.186$. Note that in this case  the fine structure of the
spectral line is not completely destroyed, as both time-scales,
$\tau_{\rm R} \approx 2826$ and $\tau_{\gamma} =2000$,  are of the
same order. In the case of strong dissipation (Fig.~\ref{tre}b), the
width of the spectral line has a Lorentzian form,
$${\rm Re}(p_{\nu}) = \gamma^2 {\rm Re}(p_0) /
(\gamma^2 + \nu^2),$$ with a width (at ${\rm Re}(p_{\nu})=1/2$)
determined by the dissipation parameter $\gamma$ ($\Delta
\nu_{\gamma} \approx 2 \gamma = 1$). The numerical results are in
good agreement, $\Delta \nu \approx 1$.  Also, the fine structure is
destroyed, as in this case $\tau_{\gamma}=2 \ll \tau_{\rm R} \approx
2826$. In \cite{Berman2004} we studied the concrete example
of the QNO interacting with an environment in which both relaxation
and decoherence take place, and we found that the frequency spectrum
behaves in a similar way as the toy model discussed in this subsection.


\subsection{Characteristic parameters for observation of quantum
effects after decoherence and relaxation}

As was discussed above,
for the simple closed quantum nonlinear system given by
Eq.~(\ref{QNO}) there are three characteristic time-scales (see
\cite{Vishik2003} for details on multi-dimensional systems). Due to
the interaction with the environment, two new time-scales appear:
$\tau_{\rm d}$ - a very short decoherence time, and $\tau_{\gamma}$
-the relaxation time. All of these five time-scales depend on the
parameters of the system and the environment. The typical region of
parameters in which one can observe quantum effects after
decoherence and relaxation is $\tau_{\rm d} \ll \tau_{\rm cl} <
\tau_{\hbar} < \tau_{\gamma} < \tau_{\rm R}$.
In the following we will consider a system which
satisfies this region of parameter a ``quasi-classical
nonlinear mesoscopic system".
The key inequality is $\tau_{\hbar} < \tau_{\gamma}$. In this case, the deviation of the quantum dynamics
from the classical one formally works as an effective ``quantum
relaxation" (or a ``quantum amplitude modulation"), which gives the
main contribution to the frequency spectral line width. The
relations between $\tau_{\rm cl}$ and $\tau_{\hbar}$ , and between
$\tau_{\rm R}$ and $\tau_{\gamma}$ are not so important. There can
be additional time-scales related to accumulation of quantum phases
\cite{Berman2004}, multi-dimensionality \cite{Vishik2003}, etc. The
details for a one-dimensional case were presented in
\cite{Dalvit2006,Berman2004}.

\section{An exact solvable example of an open quantum nonlinear system}

\subsection{Phase decoherence}

Although the PDEs described above look rather complicated,
especially for open quantum nonlinear systems, we have found the
exact solution for a quantum nonlinear oscillator interacting with the environment in the
special case of a density-density type of interaction, as in Eq.(\ref{density_density}).
The type of interaction does not provide relaxation
processes through energy exchange between the QNO and the
environment, but leads to phase decoherence. These effects result in the decay
of the amplitude of oscillations of the QNO (similar to the effects of relaxation), and
survive in the classical limit, where they correspond to the dephasing of the QNO.

We summarize here the results of \cite{Dalvit2006} in the context of the quantum-classical
transition for observables and the frequency Fourier spectrum.
We choose the system Hamiltonian $\hat{H}_{\rm S}$ as in Eq.~(\ref{QNO}),
the environment Hamiltonian $\hat{H}_{\cal E}$ as in Eq.~(\ref{qbm}), and
a density-density interaction Hamiltonian
$\hat{H}_{\rm int}$ as in Eq.~(\ref{density_density}).
For the model under consideration,  the interaction
with the environment introduces a single time scale,
$\tau_{\rm d}$ , which plays the role of a decoherence time.
This time-scale is not small, and survives even in the classical
limit: $|\alpha|^2\rightarrow\infty$, $|\beta_{\vec q}|^2\rightarrow\infty$, $\hbar\rightarrow 0$,
$\hbar|\alpha|^2=J=const$, $\hbar|\beta_{\vec q}|^2=J_{\vec
q}=const$, $\lambda_{\vec q}=const$. The typical region of
parameters in which one can observe quantum effects  is
$ \tau_{\rm cl} < \tau_{\hbar} < \tau_{\rm d}< \tau_{\rm R}$. The key inequality
is now $\tau_{\hbar} < \tau_{d}$. In this case, the deviation of the
quantum dynamics from the classical one formally works as an
effective ``quantum relaxation" (or a ``quantum amplitude
modulation"), which gives the main contribution to the frequency
spectral line width. The relations between $\tau_{\rm cl}$ and
$\tau_{\hbar}$ , and between $\tau_{\rm R}$ and $\tau_{d}$ are not
so important.

Although this model is rather trivial because the full Hamiltonian can be
diagonalized in the number basis for the joint system-environment
Hilbert space, it is a useful model system for the purposes of demonstration
of our approach. Following our previous results
\cite{Dalvit2006} it is possible to write an exact linear PDE for
any quantum dynamical observable in the joint Hilbert space
\begin{equation}
f(\alpha^*,\alpha;\beta^*_{\vec q},\beta_{\vec q};t) =
\langle \alpha, \beta_{\vec q} | \hat{f}(t) | \alpha, \beta_{\vec q}
\rangle,
\end{equation}
where $\hat{f}(t) = f(a^{\dagger}(t), a(t) ; b^{\dagger}_{\vec q}(t), b_{\vec q}(t))$ is a generic Heisenberg operator function, and
$| \alpha, \beta_{\vec q}\rangle =|\alpha\rangle\prod_j|\beta_{\vec q_j}\rangle$ is
an initial coherent state of the system and the
environment. Here $a^{\dagger}(t)$, $a(t)$, $b^{\dagger}_{\vec
q}(t)$, and $b_{\vec q}(t)$ are the Heisenberg bosonic creation and
annihilation operators for the system and the environment,
respectively, and $\vec q = (\vec q_1, \vec q_2,\ldots)$.
The corresponding PDE has the form
\begin{equation}
\frac{\partial}{\partial t} f(\alpha^*,\alpha ; \beta^*_{\vec q},
\beta_{\vec q}) = \hat{K} f(\alpha^*,\alpha;\beta^*_{\vec q},
\beta_{\vec q}; t) ,
\label{PDEopen}
\end{equation}
where the differential operator $\hat{K}$ includes the derivatives
of different orders over $\alpha^*$, $\alpha$, $\beta^*_{\vec q}$
and $\beta_{\vec q}$, and depends on the explicit form of the
corresponding full Hamiltonian.
As before, the general form of the differential operator  $\hat{K}$ is
$\hat{K} = \hat{K}_{\rm cl} + \hat{K}_{\rm q}$.
The operator  $\hat{K}_{\rm cl}$ includes only the first order derivatives
and describes the classical dynamics of the system and environment.
The operator  $\hat{K}_{\rm q}$
describes the quantum effects of the system and the environment.
The explicit expressions for both these operators are given in
\cite{Dalvit2006}.

In order to study the reduced dynamics of the system,
the function $f(\alpha^*,\alpha; \beta^*_{\vec q}, \beta_{\vec q})$
has to be traced over the variables of the environment
$\beta^*_{\vec q}, \beta_{\vec q}$. We have assumed above that
initially each environmental oscillator is populated initially in
the coherent state $|\beta_{\vec q} \rangle$.  Let us now assume
that the each environmental oscillator is initially in a mixed
thermal state at temperature $T$. Then we should perform an
additional averaging of the environmental oscillators over the
thermal distribution. The corresponding procedure is thoroughly
explained in \cite{Dalvit2006}.  The exact solution
for the system observable $\langle \alpha(\tau) \rangle_{\cal E}$, averaged
over the environmental variables, is
\begin{equation}
\langle \alpha(\tau) \rangle_{\cal E} = \alpha(\tau) R(\tau),
\label{solopen}
\end{equation}
where $\alpha(\tau)$ is defined in Eq. (3), and
\begin{eqnarray}
R(\tau) &=& \prod_{\vec q} R^{(\vec q)}(\tau), \nonumber \\
R^{(\vec q)}(\tau) &=&
\frac{1-e^{-\hbar\omega_{\vec q}/k_BT}}{1-e^{-\hbar\omega_{\vec q}/k_BT-i\hbar \lambda_{\vec q}\tau/\omega}}.
\label{Rfunction1}
\end{eqnarray}

\subsection{Classical limit}

Let us write the complex quantity $R^{(\vec q)}(\tau)$ in terms of its modulus
and phase, $R^{(\vec q)}(\tau)=e^{i\varphi^{(\vec q)}(\tau)} | R^{(\vec q)}(\tau)|$. Then,
we have from Eq.(\ref{Rfunction1})
\begin{eqnarray}
|R(\tau)| &=& e^{-\Gamma(\tau)} , \nonumber \\
\Gamma(\tau) &=& -\frac{V}{2\pi^2}\int_0^\infty dq  \,q^2\ln(| R^{(q)}(\tau)|),
\end{eqnarray}
where $V$ is the volume of the thermal bath. In the classical limit
($\hbar\rightarrow 0$) we have for $\Gamma(\tau)$
\begin{equation}
\Gamma(\tau)\approx \frac{\tau^2}{2\tau_d^2},~~\frac{1}{\tau_d^2}=
\frac{V}{2\pi^2}\frac{(k_BT)^2}{\omega^2}\int_0^\infty dq
{{q^2\lambda^2_{{\vec q}}}\over{\omega_{\vec q}^2}},
\label{Gamma}
\end{equation}
and the phase $\varphi^{({\vec q})}(\tau)$ is
\begin{equation}
\varphi^{(\vec q)}(\tau) \approx
-{{\lambda_{\vec q}k_BT}\over{\omega_{\vec q}\, \omega}}\tau.
\label{phase}
\end{equation}

The function $\alpha(\tau)$  in Eq.~(\ref{solopen}) coincides with
that in Eq.~(\ref{solutionclosed}). It is clear from
Eq.~(\ref{solopen}) that under the condition
\begin{equation}
\tau_\hbar\ll\tau_d,
\label{condition1}
\end{equation}
the width of the frequency spectrum of
$\langle \alpha(\tau) \rangle_{\cal E}$ is defined by the time-scale, $\tau_{\hbar}$, and
not by the interaction with the environment.  In the opposite case,
$\tau_\hbar\gg\tau_d$, the width of the spectral line is determined
by the interaction with the environment. A similar result was
obtained in \cite{Berman2004} for the QNO interacting via the
dipole-dipole interaction with the environment Eq.(\ref{dipole_dipole}).
But in the latter case, the time-scale $\tau_d$ in Eq.~(\ref{condition1}) should be
substituted by the relaxation time $\tau_\gamma$.


\section{Estimates for concrete systems}

Our main statement is that generally there is no classical limit for
the dynamics of quantum nonlinear systems interacting with the
environment, even when these systems are in the deep quasi-classical
region of parameters. The corresponding systems were called above
quasi-classical nonlinear mesoscopic systems (QCNMS).  In this
context we note that most classical systems surrounding us represent
a very particular exception due to (i) either an extremely deep
quasi-classicality (extremely small value of $\epsilon$) and/or (ii)
a very strong interaction with the environment. At the same time,
the general belief in the recent scientific literature is that after
the process of decoherence, the quasi-classical system can be
described by using classical probabilistic approaches. According to
the results discussed here it appears to be true only (i) for
quantum linear systems (with quadratic Hamiltonians) or (ii) for
quantum nonlinear systems with significantly small value of a
quasi-classical parameter $\epsilon$. For the QCNMS quantum effects
survive after the processes of decoherence and relaxation took
place. Moreover, these quantum effects make a crucial contribution
to the dynamics of observables. This observation may have significant
relevance for the understanding of the
properties of noise in complex quantum systems and nanodevices.
In particular, the performance of future BEC based
interferometers and nano machines will be limited by the level of
noise.

The key condition for survival of quantum effects for observables
related to the time-scale $\tau_{\hbar}$ is
$\tau_{\hbar} < \tau_{\gamma}$, which, in the simplest case of the quantum
nonlinear oscillator can be written in the form
\begin{equation}
\Theta \equiv \frac{\tau_{\gamma}}{\tau_{\hbar}} = 2 \mu_{\rm cl}
\sqrt{\epsilon} \tau_{\gamma} \gg 1 . \label{keycondition}
\end{equation}
We now present estimates for different real QCNMS that may satisfy
the above condition, and therefore may lead to the observation of
certain quantum effects that survive the process of
environment-induced decoherence and dissipation.

\subsection{Bose-Einstein condensates  in a one-dimensional toroidal geometry}

We start with a one-dimensional BEC confined in a toroidal geometry, and
described by the quantum field equation (see \cite{Berman2002,weibin}, and references therein)
\begin{equation}
i{{\partial\hat\Psi}\over{\partial
\tau}}=\Bigg[-{{\partial^2}\over{\partial\theta^2}}+2\pi\varepsilon\hat\Psi^\dagger\hat\Psi\Bigg]\hat\Psi.
 \label{eq1}
\end{equation}
Here $\varepsilon=4 R a / S$, $R$ is the radius of the toroidal trap, $S$ is
the area of the cross-section of the torus, and
$a$ the interatomic s-wave scattering length ($a>0$ for a repulsive
interaction, and $a<0$ for an attractive interaction).
The dimensionless time is  $\tau=\hbar t / 2 m R^2$.
The operator  $\hat\Psi(\theta,\tau)$ can be expanded as
\begin{equation}
\hat\Psi(\theta,\tau)={{1}\over{\sqrt{2\pi}}}\sum_{k=-\infty}^{\infty}\hat a_k(\tau)e^{ik\theta}.
 \label{eq3}
\end{equation}
Here $\hat a_k(\tau)$ and $\hat a_k^\dagger(\tau)$ are
annihilation and creation bosonic operators, respectively, and
the field operator is periodic
$\hat\Psi(\theta+2\pi,\tau)= \hat\Psi(\theta,\tau)$ and satisfies the normalization
condition
\begin{equation}
\int_0^{2\pi}\hat\Psi^\dagger(\theta,\tau)\hat\Psi(\theta,\tau)d\theta=\sum_{k=-\infty}^{\infty}\hat
n_k\equiv\hat N,
\label{eq4}
\end{equation}
where $\hat n_k$ is the operator of the number of particles in the
mode with momentum $k$, and $\hat N$ is the operator of the total
number of particles.

In the following we only consider the case of repulsive interactions, $a>0$.
>From Eqs.~(\ref{eq1}) and (\ref{eq3})  it follows that the operators $\hat a_k(\tau)$
satisfy the following system of coupled first-order differential equations:
\begin{equation}
i\dot {\hat a}_k= k^{2} \hat a_{k}+
\varepsilon\sum_{k_1,k_2,k_3=-\infty}^\infty \hat a_{k_1}^\dag \hat a_{k_2} \hat a_{k_3}
\delta_{k+k_1-k_2-k_3,0},
\label{eq.qmot}
\end{equation}
where ``dot" means derivative with respect to $\tau$.
Here we shall limit ourselves to consider only a single mode in
Eq.(\ref{eq.qmot}): $\hat a_k^\prime=\hat a_k\delta_{k^\prime,k}$,
which is stable under the condition $a>0$ (For a more general case
see (\cite{bikt1,bt})). In this simplified case Eq. (\ref{eq.qmot}) takes the
form
\begin{equation}
i\dot {\hat a}_k  = k^{2} \hat a_{k} + \varepsilon \hat a_{k}^\dag \hat a^2_{k}=[\hat a_k,\hat H_{\rm eff}],
\label{eq.onemode}
\end{equation}
with the effective Hamiltonian
\begin{equation}
\hat H_{\rm eff}=\bigg(k^2-{{\varepsilon}\over{2}}\bigg)\hat a^\dagger_k\hat
a_k+{{\varepsilon}\over{2}}(\hat a^\dagger_k\hat a_k)^2.
\label{eq.onemodehamiltonian}
\end{equation}

To solve the system Eqs.(\ref{eq.onemode}), (\ref{eq.onemodehamiltonian})
we use the above described techniques of projection onto the basis
of coherent states. Let us assume that at $\tau=0$ the $k$th mode of the
bosonic field can be represented by a coherent state,
$|\alpha_k\rangle$, described by a complex number $\alpha_{k}$. We
denote
\begin{equation}
\alpha_{k} (\tau)= \langle {\alpha_k} |\, \hat a_{k} (\tau)\, |{\alpha_k} \rangle
= \alpha_{k} (t, {\alpha_k}, {\alpha_k}^{\ast}).
\end{equation}
Note that all atoms $N$ occupy the single mode $k$, that is
$\langle\alpha_k|\hat n_k|\alpha_k\rangle\equiv n_k=N$.
The exact linear PDE for the observable $\alpha_k(\tau)$ is
\begin{equation}
\begin{array}{ll}
\dot{\alpha}_{k} (\tau)=\hat{K}\, \alpha_{k} (\tau),\\
\alpha_{k} (0)= \alpha_{k},
\end{array}
\label{eq.HeisP}
\end{equation}
where
\begin{eqnarray}
\hat{K} &=& i\bigg(k^2+\varepsilon|\alpha_k|^2\bigg)
\Big(  \alpha^*_{k} \frac{\partial}{\partial \alpha^*_{k}} - c.c. \Big) \nonumber \\
&&  +   i\frac{\varepsilon}{2} \Big( (\alpha^*_k)^2 \frac{\partial^2}{\partial (\alpha^*_k)^2} - c.c. \Big) .
\end{eqnarray}
It is more convenient to write these equations using action-angle variables. Namely,
instead of the variables $\alpha_k$ and $\alpha^*_k$ we
use the variables $n_k$ (remember that in this simplified case the number of
atoms in mode $k$ is fixed, $n_k=N$) and
$\theta_k$, where
\begin{equation}
\alpha_k=\sqrt{n_k}e^{-i\theta_k}.
\end{equation}
Using the expressions
$$
\Big( \alpha^*_{k} \frac{\partial}{\partial \alpha^*_{k}} - c.c \Big)=-i{{\partial}\over{\partial\theta_k}},
$$
$$
\Big( (\alpha^*_k)^2 \frac{\partial^2}{\partial (\alpha^*_k)^2} - c.c. \Big)=
i{{\partial}\over{\partial\theta_k}}-2iN{{\partial^2}\over{\partial N\partial\theta_k}},
$$
one can derive the following equation for
$\alpha_k(\tau)$ in new variables
\begin{equation}
{{\partial\alpha_k}(\tau)\over{\partial\tau}}=\bigg(k^2-{{\varepsilon}\over{2}}+\varepsilon
N\bigg){{\partial\alpha_k}(\tau)\over{\partial\theta_k}}+\varepsilon
N{{\partial^2\alpha_k}(\tau)\over{\partial N\partial\theta}}.
\label{eq.newvariables}
\end{equation}
This equation possesses a
solution of the form of a finite amplitude periodic wave
\begin{equation}
\alpha_{k}(\tau)=\exp\Big\{-ik^2\tau-(1 -
\exp (-i\varepsilon\tau)) |\alpha_{k}|^{2} \Big\}\alpha_{k} .
\label{eq.fapws}
\end{equation}
This solution
has two characteristic time-scales
\begin{equation}
\tau_\hbar={{1}\over{|\alpha_k||\varepsilon|}},~
\tau_R={{2\pi}\over{|\varepsilon|}} .
 \label{times}
\end{equation}
The first one describes the breakdown of quantum-classical correspondence,
and the second one is the time-scale of quantum revivals
\cite{BermIomiZasl1,Berman2002,Pit1}.
Note that Eq. (\ref{eq.fapws}) formally turns into the GP
solution (which we also will call a ``classical" field theory solution)
\begin{equation}
\alpha_{k}^{cl}(\tau)=\exp\Big\{-i\Big(k^2+\varepsilon
|\alpha_{k}|^{2}\Big)\tau
\Big\}\alpha_{k},
\label{cl}
\end{equation}
when $|\varepsilon|\rightarrow 0$,
$|\alpha_k|^2=N\rightarrow\infty$, and
$|\varepsilon||\alpha_k|^2=const$.

For this one-dimensional BEC system the condition  Eq.(\ref{keycondition})
for observation of quantum effects after decoherence and relaxation is reduced to the
following
\begin{equation}
t_\hbar =\frac{mRS}{{2}\hbar|\alpha|a}\ll t_\gamma,
\label{condition}
\end{equation}
where $t_\gamma$ is the relaxation time, and $m$ is the mass of the BEC atom.
To estimate the time-scale $t_\hbar$ we assume that
$N=10^3$ $^{87}$Rb atoms ($a=2.5\times 10^{-6} {\rm cm}$) are trapped in
a toroidal trap with radius $R=5\times 10^{-4} {\rm cm}$ and cross-section
$S=10^{-8} {\rm cm}^2$, which implies $t_\hbar\sim 4.5$ms. The
corresponding bandwidth of the frequency spectrum, which
characterizes the quantum effects related to the time-scale
$t_\hbar$, is $\Delta \nu\approx 2\sqrt{2}/t_\hbar\approx 0.6$kHz.

\subsection{Relation between the time-scale $t_\hbar$ and phase diffusion of two
Bose-Einstein condensates}

The effect of phase diffusion of the relative phase between two BECs due to
atomic collisions was studied theoretically in  \cite{phase_diffusion}
and was recently observed in the high atomic density regime with two BECs
trapped on an atom chip \cite{Ketterle2007}.

Let us summarize here the main ideas behind phase diffusion in the simplest ideal case.
Imagine a Bose-Einstein condensate that is symmetrically split into two pieces via a double-well potential. Assuming that the split process is slow enough (i.e., the barrier is raised on a time scale long compared
to the inverses of the excitation frequencies of the initial potential well), but fast enough to freeze
the relative phase between the two condensates in each well, the final state of the condensates
after the split can be described as a state $|\varphi \rangle$ that is a superposition over many relative
number states
\begin{equation}
|\varphi \rangle = \sum_{k=0}^{N} e^{i \varphi k} \, \sqrt{\frac{N!}{2^N k! (N-k)!}} \, |k, N-k\rangle,
\label{state_split}
\end{equation}
where $N$ is the total number of atoms and $\varphi$ is the relative phase between the condensates
in each well. Here we have assumed that during the split process the atomic interactions are
negligible. For simplicity we will also assume that the relative phase between the two condensates
is zero, $\varphi=0$. After the split, each condensate evolves independently (the barrier is sufficiently raised
to suppress tunneling between the wells). Because of atom-atom interactions, the energy of number
states $E(k,N-k)$ have a quadratic dependence on the atom numbers in each well, $k$ and $N-k$,
so that the different relative number states have different phase evolution rates. The state vector
(\ref{state_split}) evolves as
\begin{equation}
|\chi, t \rangle = e^{-i \omega t} \sum_{k=0}^N \sqrt{\frac{N!}{2^N k! (N-k)!}} \,
e^{\xi (k-N/2)^2 } \, |k, N-k \rangle ,
\label{state_evolved}
\end{equation}
where $\omega=2 E(N/2)/\hbar$ is the frequency of each well, and
$\xi =\frac{1}{\hbar} \, \frac{d^2E(k)}{dk^2}|_{k=N/2}=2 \mu \hbar$ is the effect of nonlinearities.
To study the phase distribution of the evolved state one projects this evolved state onto phase states. These are orthonormal
states of the form
\begin{equation}
|\phi_p \rangle = \frac{1}{\sqrt{N=1}} \sum_{k=0}^N e^{i \phi_p} \, |k, N-k \rangle ,
\end{equation}
with $\phi_p = 2 \pi p/ (N+1)$ and $p=-N/2, \ldots, N/2$. In the limit $N\gg 1$, the phase distribution
of the state (\ref{state_evolved}) is
$P(\phi)=|\langle \phi | \chi,t\rangle|^2 = \sqrt{\pi / 2 (\Delta \phi)^2} \, \exp(-\phi^2/ 2 (\Delta \phi)^2)$,
with a phase dispersion that evolves in time as
\begin{equation}
\Delta \phi(t)^2 = \Delta \phi_0^2 + R^2 t^2 ,
\label{phase_dispersion}
\end{equation}
where $\Delta \phi_0^2=1/N$ is the phase dispersion for the initial two-model coherent
state (\ref{state_split}), and $R=\sqrt{N} \xi$ is the rate of phase diffusion. This rate defines
a phase diffusion time-scale
\begin{equation}
t_{\rm ph.diff}=  \frac{1}{2 \bar{\mu} \sqrt{N} } ,
\label{time_diffusion}
\end{equation}
which coincides with the time-scale $\tau_{\hbar}$ in Eq.(\ref{time_hbar}) of breakdown of
quantum-classical correspondence of the quantum nonlinear oscillator initially prepared in
a coherent state with mean number of excitations $N=|\alpha|^2$.

We conclude from the above considerations that an alternative way to observe the effect of
the ``quantum" time scale $\tau_{\hbar}$ in the dynamics of the quantum nonlinear oscillator (QNO)
is to analyze phase diffusion of two condensates (which can be modeled as two uncoupled
QNOs after the splitting process), initially prepared in a quasi-classical coherent state,
as in Eq.(\ref{state_split}). A related experiment was performed in \cite{Ketterle2007}
for an initial two-mode number-squeezed state instead of a two-mode coherent state.
In that case, the initial phase dispersion is much wider, $\Delta \phi_0^2 \simeq s / N$,
and the rate of phase diffusion is much larger, $R \simeq s \sqrt{N} \xi$, where $s \simeq \sqrt{N} \gg 1$ is the squeezing parameter.

\subsection{The time-scale $\tau_\hbar$ for mechanical resonators and for nonlinear optical systems}

For a mechanical resonator or cantilever
the quasi-classical parameter is $\epsilon = 1/n$,  where $n$ is the
average number of levels involved in the quantum state of the resonator, that we assume to be a coherent state.  The dimensionless relaxation time is
$\tau_{\gamma} = 2 Q$, where $Q$ is the resonator's quality factor.
The condition Eq.~(\ref{keycondition}) takes the form
\begin{equation}
\Theta_{\rm cantilever} = \frac{4 \mu_{\rm cl} Q}{\sqrt{n}} \gg 1.
\label{cantilever}
\end{equation}
Different aspects of cantilevers, from kilohertz to gigahertz
frequencies, including their nonlinear properties, are discussed,
for example, in \cite{Stipe2001,Robert2005}.

A condition similar to Eq.~(\ref{cantilever}) holds for quantum
nonlinear optical systems in high quality resonators. In this case,
$n$ is the average number of photons
in the initially coherent state of the cavity resonance  mode, and the
classical parameter of nonlinearity can be written as
$\mu_{\rm cl} = \chi J / \omega_{\rm cav}$, where $\chi$  is the
nonlinear susceptibility, and $\omega_{\rm cav}$  is the cavity
resonance frequency \cite{Walls1994}.


\section{Conclusions}

In this paper we have reviewed the effects of singular perturbations
resulting from quantum terms in the dynamical equations for
observables of open quantum nonlinear quasi-classical systems. We
have argued that when the time-scale for quantum-classical
departure, generally given by the  time-scale $\tau_{\hbar}$, is
much shorter than the dissipation time scale $\tau_{\gamma}$,
certain quantum effects survive the process of decoherence, and
could be observed from characteristic properties of the
time-evolution of observables, such as in the frequency spectrum and
in the noise spectrum. With recent advances in quantum technology we
expect that the key condition (\ref{keycondition}) for detecting
such effects may be experimentally realized, and quantum effects
related to the time-scale $\tau_\hbar$ can be observed in the
quasi-classical region of parameters.

\section{Acknowledgments}

We are thankful to M.G. Boshier,  B.M. Chernobrod, L. Pezz\'{e}, and
E.M. Timmermans for useful discussions. Part of this work was done
during the  stay of GPB and DARD at the Institut Henri
Poincare-Centre Emile Borel. The authors thank this institution for
hospitality and support. This work was carried out under the
auspices of the National Nuclear Security  Administration of the
U.S. Department of Energy at Los Alamos National Laboratory under
Contract No. DE-AC52-06NA25396.

\end{document}